# LIMITING SELF-PROPAGATING MALWARE BASED ON CONNECTION FAILURE BEHAVIOR THROUGH HYPER-COMPACT ESTIMATORS

You Zhou[1], Yian Zhou[1], Shigang Chen[1] and O. Patrick Kreidl[2]

[1]Department of Computer & Information Science & Engineering, University of Florida,
Gainesville, FL, USA 32611
`youzhou,yian,sgchen@cise.ufl.edu`
[2]Department of Electrical Engineering, University of North Florida,
Jacksonville, FL, USA 32224
`patrick.kreidl@unf.edu`

## ABSTRACT

*Self-propagating malware (e.g., an Internet worm) exploits security loopholes in software to infect servers and then use them to scan the Internet for more vulnerable servers. While the mechanisms of worm infection and their propagation models are well understood, defense against worms remains an open problem. One branch of defense research investigates the behavioral difference between worm-infected hosts and normal hosts to set them apart. One particular observation is that a worm-infected host, which scans the Internet with randomly selected addresses, has a much higher connection-failure rate than a normal host. Rate-limit algorithms have been proposed to control the spread of worms by traffic shaping based on connection failure rate. However, these rate-limit algorithms can work properly only if it is possible to measure failure rates of individual hosts efficiently and accurately. This paper points out a serious problem in the prior method. To address this problem, we first propose a solution based on a highly efficient double-bitmap data structure, which places only a small memory footprint on the routers, while providing good measurement of connection failure rates whose accuracy can be tuned by system parameters. Furthermore, we propose another solution based on shared register array data structure, achieving better memory efficiency and much larger estimation range than our double-bitmap solution.*

## 1. INTRODUCTION

Self-propagating malware (e.g., an Internet worm) exploits security loopholes in server software. It infects vulnerable servers and then uses them to scan the Internet for more vulnerable servers [1 - 3]. In the past two decades, we have witnessed a continuous stream of new worms raging across the Internet [4 - 7], sometimes infecting tens of thousands or even millions of computers and causing widespread service disruption or network congestion. The mechanisms of worm propagation have been well understood [8 - 11], and various propagation models were developed [12 - 15] to demonstrate analytically the properties of how worms spread among hosts across networks. Significant efforts have also been made to mitigate worms, with varying degrees of success and limitations. Worms remain a serious threat to the Internet.

Patching defects in software is the most common defense measure, not only to worms but also to other types of malware. However, it is a race for who (good guys or bad guys) will find the security defects first. Software is vulnerable and its hosts are subject to infection before the security problems are identified and patched. Moreover, not all users will patch their systems timely, leaving a window of vulnerability to the adversary that will try to exploit every opportunity. Moore et al. investigated worm containment technologies such as address blacklisting and content filtering, and such systems must interdict nearly all Internet paths in order to be successful [13]. Williamson proposed to modify the network stack to bound the rate of connection requests made to distinct destinations [16]. To be effective, it requires a majority of all Internet hosts are upgraded to the new network stack, which is difficult to realize. Similar Internet-wide upgrades are assumed by other host-based solutions in the literature, each employing intrusion detection and automatic control techniques whose supporting models must be calibrated for the specific machine that they will reside upon [17 - 20].

Avoiding the requirement of coordinated effort across the whole Internet, the distributed anti-worm architecture (DAW) [21] was designed for deployment on the edge routers of an Internet service provider (ISP) under a single administrative control. DAW observes a behavioral difference between worm-infected hosts and normal hosts: as an infected host scans random addresses for vulnerable hosts, it makes connection attempts but most will fail, whereas normal users's connection attempts to their familiar servers are mostly successful. By observing the failed connections made by the hosts, the edge routers are able to separate out hosts with large failure rates and contain the propagation of the worms. With a basic rate-limit algorithm, a temporal rate-limit algorithm and a spatial rate-limit algorithm, DAW offers the flexibility of tightly restricting the worm's scanning activity, while allowing the normal hosts to make successful connections at any rate.

However, for rate limit to work properly, we must be able to measure the connection failure rates of individual hosts accurately and efficiently. This paper points out that using Internet Control Message Protocol (ICMP) messages for this purpose [21] is flawed as they are widely blocked on today's Internet, and the total number of message packets in this big data [22] [23] [30] [31] and cloud computing [36 - 39] era is enormous. In this paper, we first design a novel measurement method that solves the problem with a highly efficient data structure based on bitmaps [41], which keeps record of connection attempts and results (success or fail) in bits, from which we can recover the connection failure rates, while removing the duplicate connection failures (which may cause bias against normal hosts). Our double-bitmap solution is highly efficient for online per-packet operations, and the simulation results show that not only does the data structure place a small memory footprint on the routers, but also it provides good measurement of connection failure rates whose accuracy can be tuned by system parameters. However, we discover the double-bitmap solution has a limitation: it is difficult to extend the estimation range in such memory constrain without causing estimation inaccuracy. In order to address this problem, we propose another solution based on shared register array data structure [42], achieving better memory efficiency and much larger estimation range than our double-bitmap solution.

The rest of the paper is organized as follows: Section 2 gives the propagation model of random-scanning worms and reviews the rate-limit algorithms based on connection failure rates. Section 3 explains the problem causing inaccurate failure rate measurement and provides a novel solution with double bitmaps. Section 4 describes our shared register array solution in detail. Section 5 presents simulation results. Section 6 draws the conclusion.

## 2. BACKGROUND

### 2.1. Propagation of Random-scanning Worms

This paper considers a type of common worms that replicates through random scanning of the Internet for vulnerable hosts. Their propagation can be roughly characterized by the classical simple epidemic model [26 - 28]:

$$\frac{di(t)}{d(t)} = \beta i(t)(1-i(t)), \quad (1)$$

where $i(t)$ is the percentage of vulnerable hosts that are infected with respect to time $t$, and $\beta$ is the rate at which a worm-infected host detects other vulnerable hosts. More specifically, it has been derived [27] that the derivative formula of worm propagation is

$$\frac{di(t)}{dt} = r\frac{V}{N}i(t)(1-i(t)), \quad (2)$$

where $r$ is the rate at which an infected host scans the address space, $N$ is the size of the address space, and $V$ is the total number of vulnerable hosts.

Solving the equation, the percentage of vulnerable hosts that are infected over time is

$$i(t) = \frac{e^{r\frac{V}{N}(t-T)}}{1+e^{r\frac{V}{N}(t-T)}}.$$

Let $v$ be the number of initially infected hosts at time 0. Because $i(0) = v/V$, $T = -\frac{N}{r \cdot V}\ln\frac{v}{V-v}$. Solving this logistic growth equation for $t$, we know the time it takes for a percentage $\alpha(\geq v/V)$ of all vulnerable hosts to be infected is

$$t(\alpha) = \frac{N}{r \cdot V}(\ln\frac{\alpha}{1-\alpha} - \ln(\frac{v}{V-v})). \quad (3)$$

It is clear that $t(\alpha)$ is inversely proportional to the scanning rate $r$, which is the number of random addresses that an infected host attempts to contact (for finding and then infecting vulnerable hosts) in a certain measurement period. If we can limit the rate of worm scanning, we can slow down their propagation, buying time for system administrators across the Internet to take actions.

## 2.2. Behavior-based Rate-Limit Algorithms

In order to perform rate-limit, we need to identify hosts that are likely to be worm-infected. One way to do so is observing different behaviors exhibited from infected hosts and normal hosts. One important behavioral observation was made by [21], which argues that infected hosts have much larger failure rates in their initiated Transfer Contorl Protocol (TCP) connections than normal hosts. We can then apply rate limits to hosts with connection failure rates beyond a threshold and thus restrict the speed at which worms are spread to other vulnerable hosts. (Same as our work, the paper [21] studies worms that spread via TCP, which accounts for the majority of Internet traffic.) Below we briefly describe the host behavior difference in connection failure rate, which is defined as the number of failed TCP connection attempts made by a source host during a certain measurement period, where each attempt corresponds to a SYN packet and each SYN-ACK signals a successful attempt, while the absence of a SYN-ACK means a failure.

- Suppose a worm is designed to attack a software vulnerability in a certain version of web servers from a certain vendor. Consider an arbitrary infected host. Let $N$ be the total number of possible IP addresses and $N'$ be the number of addresses held by web servers, which listen to port 80. $N' \ll N$ because web servers only account for a small fraction of the accessible Internet. As the infected host picks a random IP address and sends a SYN packet to initiate a TCP connection to port 80 of that address, the connection only has a chance of $N'/N$ to be successful. It has a chance of $1 - N'/N \approx 1$ to fail. The experiment in [21] shows that only 0.4% of all connections made to random addresses at TCP port 80 are successful. Together with a high scanning rate, the connection failure rate of an infect host will be high. Moreover, the measured connection failure rate is an approximation of the host's scanning rate.

- The connection failure rate of a normal host is generally low because a typical user accesses pre-configured servers (such as mail server and DNS server) that are known to be up for most of the time. An exception is web browsing, where the domain names of web servers are used, which again lead to successful connections for most of the time according to our experiences. Cases when the domain names are mistyped, it result in DNS lookup failure and no connection attempts will be made --- consequently no connection failure will occur.

By measuring the connection failure rates of individual hosts, the paper [21] proposes to limit the rate at which connection attempts are made by any host whose failure rate exceeds a certain threshold. By limiting the rate of connection attempts, it reduces the host's connection failure rate back under the threshold. An array of rate-limit algorithms were proposed. The basic algorithm rate-limits individual hosts with excessive failure rates. The temporal rate-limit algorithm can tolerate temporary high failure rates of normal hosts but make sure the long-term average failure rates are kept low. The spatial rate-limit algorithm can tolerate some hosts' high failure rates but make sure that the average failure rates in a network are kept low.

An important component that complements the rate-limit algorithms is the measurement of connection failure rates of individual hosts. This component is however not adequately addressed by [21]. As we will point out in the next section, its simple method does not provide accurate measurement on today's Internet. We will provide two new methods that can efficiently solve this important problem with two novel data structures: bitmap and shared register array.

## 3. A DOUBLE-BITMAP SOLUTION FOR LIMITING WORM PROPAGATION

In this section, we explain the problem that causes inaccurate measurement of connection failure rates and provide a new measurement solution that can work with existing rate-limit algorithms to limit worm propagation.

### 3.1. Failure Replies and the Problem of Blocked ICMP Messages

We first review the method of measuring the connection failure rates in [21]. After a source host sends a SYN packet to a destination host, the connection request fails if the destination host does not exist or does not listen on the port that the SYN is sent to. In the former case, an ICMP host-unreachable packet is returned to the source host; in the latter case, a TCP RESET packet is returned. The ICMP host-unreachable or TCP RESET packet is defined as a *failure reply*. The connection failure rate of a host $s$ is measured as the rate of failure replies that are sent to $s$. The rationale behind this method [21] is that the rate of failure replies sent back to the source host should be close to the rate of failed connections initiated by the host. The underlying assumption is that, for each failed connection, a failure reply (either an ICMP host-unreachable packet or a TCP RESET packet) is for sure to be sent back to the source host.

However, this assumption may not be realistic. Today, many firewalls and domain gateways are configured to suppress failure replies. In particular, many organizations block outbound ICMP host-unreachable packets because attacks routinely use ICMP as a reconnaissance tool. When the ICMP host-unreachable packets are blocked, the rate of failure replies sent back to a source host will be essentially much lower than the rate of failed connections that the host has initiated. In other words, a potential worm-affected host may initiate many failed connections, but only a handful of failure replies will be sent back to it. Under these circumstances, the connection failure rate measured by failure replies will be far lower than the actual failure rate, which in turn misleads the rate-limit algorithms and makes them less effective.

To make the problem more complicated, when we measure the connection failure rates of individual hosts, all failed connections made from the same source host to the same destination host in each measurement period should be treated as duplicates and thus counted only once. We use an example to illustrate the reason: Suppose the mail server of a host is down and the email reader is configured to automatically attempt to connect to the server after each timeout period (e.g., one minute). In this case, a normal host will generate a lot of failed connections to the same destination, pushing its connection failure rate much higher than the usual value (when the server is not down) and falsely triggering the rate-limit algorithms to restrict the host's access to the Internet. Therefore, when we measure the connection failure rate of a source host, we want to remove the duplicates to the same destination and measure the rate of failed connections to distinct destinations.

## 3.2. SYN/SYN-ACK Solution and Problems of Duplicate Failures and Memory Consumption

We cannot use failure replies to measure the connection failure rates. Another simple solution is to use SYN and SYN-ACK packets. Each TCP connection begins with a SYN packet from the source host. If a SYN-ACK packet is received, we count the connection as a successful one; otherwise, we count it as a failed connection. (Technically speaking, a third packet of ACK from the source to the destination completes the establishing of the connection. For our anti-worm purpose, however, the returned SYN-ACK already shows that the destination host is reachable and listens to the port, which thus does not signal worm behavior --- random scanning likely hits unreachable hosts or hosts not listening to the port.)

Using SYN and SYN-ACK packets, a naive solution is for each edge router to maintain two counters, $k_s$ and $k_r$, for each encountered source address, where $k_s$ is the rate of SYN packets sent by the source (i.e., the number of SYN packets sent during a measurement period), and $k_r$ is the rate of SYN-ACK packets received by the source (i.e., the number of SYN-ACKs received during a measurement period). The connection failure rate $k$ is simply $k_s - k_r$.

This simple solution is memory efficient, as it only requires 64 bits per source host for failure rate measurement, assuming each counter takes 32 bits. However, this solution cannot address the problem of duplicate failures. As discussed in Section 3.1, when we measure the connection failure rate of a source host, we want to remove the duplicates to the same destination in the same measurement period, because measuring duplicate failures may cause bias against normal hosts. Maintaining two counters alone cannot achieve the goal of removing duplicate failures.

An alternative solution is to have the edge router store a list of distinct destination addresses for each source host. However, such per-source information consumes a large amount of memory. Suppose each address costs 32 bits. The memory required to store each source host's address list will grow linearly with the rate of distinct destination hosts that the source host initiates connection requests to. For example, the main gateway at our campus observes an average of more than 10 million distinct source-destination pairs per day. If the edge router keeps per-source address list, it will cost more than 320 megabits of memory, which soon exhausts the small on-die SRAM memory space of the edge router. Therefore, this solution is not feasible either.

The major goal of this paper is to accurately measure the connection failure rates with a small memory. However, tradeoffs must be made between measurement accuracy and memory consumption under the requirement of duplicate failure removal. Existing research uncovered the advantages of using Bloom filters [28] [29] or bitmaps [24] [25] [32 - 35] [40] to compress the connection information in limited memory space and automatically filter duplicates, which can be adopted to measure the connection failure rates. For example, the edge router can maintain two bitmaps for each source host, and map each SYN/SYN-ACK packet of the host into a bit in the host's corresponding bitmap, from which the rate of SYN/SYN-ACK packets of each host can be recovered. However, the measurement accuracy depends on setting the bitmap size for each source host properly in advance. In practice, it is difficult to pre-determine the values as different source hosts may initiate connection requests at unpredictable and different rates, which limits the practicability of this solution as well.

### 3.3. Double Bitmaps

In order to address the problems of duplicate failures and memory consumption, instead of using per-source address lists or bitmaps, we incorporate two shared bitmaps to store the SYN/SYN-ACK information of all source hosts. Our double-bitmap solution includes two phases: in the first phase, the edge router keeps recoding the SYN/SYN-ACK packets of all source hosts through setting bits in the bitmaps; in the second phase, the network management center will recover the connection failure rates from the two bitmaps based on maximum likelihood estimation (MLE), and notify the edge router to apply rate limit algorithms to limit the connection attempts made by any host whose failure rate exceeds some threshold. Below we will explain the two phases, and then mathematically derive an estimator to calculate the connection failure rate.

#### 3.3.1. Phase I: SYN / SYN-ACK Encoding

In our solution, each edge router maintains two bitmaps $B_s$ and $B_r$, which encode the distinct SYN packets and SYN-ACK packets of all source hosts within its network, respectively. Let $m_s$ and $m_r$ be the number of bits in $B_s$ and $B_r$ correspondingly. Below we will explain how an edge router encodes the distinct SYN packet information into $B_s$, which can later be used to estimate the SYN sending rate $k_s$ for each source host. The way for the edge router to encode the distinct SYN-ACK packet information into $B_r$ is quite similar, which we omit.

For each source host $src$, the edge router randomly selects $l_s$ ($\ll m_s$) bits from the bitmap $B_s$ to form a logical bitmap $src$, which is denoted as $LB(src)$. The indices of the selected bits are $H(src \oplus R[0])$, $H(src \oplus R[1])$, $\cdots$, $H(src \oplus R[l_s - 1])$, where $\oplus$ is bitwise XOR, $H(\cdots)$ is a hash function whose range is $[0, m_s)$, and $R$ is an integer array storing randomly chosen constants to arbitrarily alter the hash result. Similarly, the logical bitmap can be constructed from $B_s$ for any other hosts. Essentially, we embed the bitmaps of all possible hosts in $B_s$. The bit-sharing relationship is dynamically determined on the fly as each new host $src'$ will be allocated a logical bitmap $LB(src')$ from $B_s$ to store its SYN packet information.

Given above notations and data structures, the online coding works as follows. At the beginning of each measurement period, all bits in $B_s$ are reset to zeros. Suppose a SYN packet signatured with a $\langle src, dst \rangle$ host address pair is routed by the edge router. The router will randomly select a bit from the logical bitmap $LB(src)$ based on $src$ and $dst$, and set this bit in $B_s$ to be one. The index of the bit to be set for this SYN packet is given as follows:

$$H(src \oplus R[H(dst \oplus K) \bmod l_s]).$$

The second hash, $H(dst \oplus K)$, ensures that the bit is pseudo-randomly selected from $LB(src)$, and the private key $K$ is introduced to prevent the hash collision attacks. Therefore, the overall effect to store the SYN packet information is:

$$B[H(src \oplus R[H(dst \oplus K) \bmod l_s])] = 1.$$

Similarly, the edge router only needs to set a bit in the bitmap $B_r$ to be one for each SYN-ACK packet using the same mechanism. Note that in our solution, to store a SYN/SYN-ACK packet, the router only performs two hash operations and sets a single bit in its bitmap, which is quite efficient. In addition, duplicates of SYN or SYN-ACK information with same $\langle src, dst \rangle$ signature will mark the same bit in the shared bitmaps such that the duplicate information is filtered as desired.

### 3.3.2. Phase II: Failure Rate Measurement

At the end of each measurement period, the edge router will send the two bitmaps $B_s$ and $B_r$ to the network management center (NMC), which will estimate connection failure rate $k$ for each source host $src$ based on $B_s$ and $B_r$, and notify the edge router to apply rate limit algorithms to limit the connection attempts made by any host whose failure rate exceeds some threshold. Since rate-limit algorithms have been fully studied in [21], we will focus on the measurement of connection failure rates based on the register arrays. The measurement process is described in the following.

First, the NMC extracts the logical bitmaps $LB(src)$ and $LB'(src)$ of each source host $src$ from the two bitmaps $B_s$ and $B_r$, respectively. Second, the NMC counts the number of zeros in $LB(src)$, $LB'(src)$, $B_s$ and $B_r$, which are denoted by $U_s^l$, $U_r^l$, $U_s^m$, and $U_r^m$, respectively. Then the NMC divides them by the corresponding bitmap size $l_s$, $l_r$, $m_s$, and $m_r$, and calculates the fraction of bits whose values are zeros in $LB(src)$, $LB'(src)$, $B_s$ and $B_r$ correspondingly. That is, $V_s^l = U_s^l / l_s$, $V_r^l = U_r^l / l_r$, $V_s^m = U_s^m / m_s$, and $V_r^m = U_r^m / m_r$. Finally, the NMC uses the following formula to estimate connection failure rate $k$ for source host $src$:

$$\hat{k} = \frac{\ln V_s^l - \ln V_s^m}{\ln(1 - \frac{1}{l_s}) - \ln(1 - \frac{1}{m_s})} - \frac{\ln V_r^l - \ln V_r^m}{\ln(1 - \frac{1}{l_r}) - \ln(1 - \frac{1}{m_r})} \tag{4}$$

### 3.3.3. Derivation of the estimator

Now we follow the standard MLE method to get the MLE estimators $\hat{k}_s$ and $\hat{k}_r$ of $k_s$ and $k_r$, respectively, and then derive $\hat{k}$ given by (4). Since the way to derive the MLE estimator for $k_s$ and $k_r$ is quite similar, we will only derive the MLE estimator formula for $\hat{k}_s$, and directly give the result for $\hat{k}_r$. To derive $\hat{k}_s$, we first analyze the probability $q(k_s)$ for an arbitrary bit in $LB(src)$ to be '0', and use $q(k_s)$ to establish the likelihood function $L$ to observer $U_s^l$ '0' bits in $LB(src)$. Finally, maximizing $L$ with respect to $k_s$ will lead to the MLE estimator, $\hat{k}_s$.

Note that $k_s$ is the actual rate of distinct SYN packets sent by a source host $src$, and $n_s$ is the rate of distinct SYN packets sent by all hosts within the router's network. Consider an arbitrary bit $b$ in $LB(src)$. A SYN packet sent by $src$ has a probability of $1/l_s$ to set $b$ to '1', and a SYN packet sent by any other host has a probability of $1/m_s$ to set $b$ to '1'. Hence, the probability $q(k_s)$ for bit $b$ to remain '0' at the end of the measurement period is

$$q(k_s) = \left(1 - \frac{1}{m_s}\right)^{n_s - k_s} \left(1 - \frac{1}{l_s}\right)^{k_s}. \tag{5}$$

Because the bits in any logical bit array are randomly selected from the bitmap $B_s$, each of the $n_s$ SYN packets has about the same probability of $1/m_s$ to choose any bit in $B_s$. So for an arbitrary bit in $B_s$, the probability for it to be '0' after storing all $n_s$ distinct SYN packets is

$$q(n_s) = \left(1 - \frac{1}{m_s}\right)^{n_s}. \tag{6}$$

In this sense, the number of zero bits in $B_s$ follows a binomial distribution $U_s^m \sim B(m_s, q(n_s)) = B(m_s, (1 - 1/m_s)^{n_s})$. Therefore, the expected value for $V_s^m$ is

$$E(V_s^m) = E\left(\frac{U_s^m}{m_s}\right) = \frac{m_s(1 - \frac{1}{m_s})^{n_s}}{m_s} = q(n_s). \tag{7}$$

Substituting (7) to (5), and replacing $E(V_s^m)$ by its instance value $V_s^m$, we have the following instance value for $q(k_s)$:

$$q(k_s) = V_s^m \times \left(\frac{1 - 1/l_s}{1 - 1/m_s}\right)^{k_s}. \tag{8}$$

Given the probability for each bit in $LB(src)$ to be '0' as $q(k_s)$, we can establish the likelihood function to observe $U_s^l$ '0' bits in $LB(src)$ as follows:

$$L = q(k_s)^{U_s^l}(1-q(k_s))^{l_s-U_s^l}. \tag{9}$$

The MLE estimator of $k_s$ is the value of $k_s$ that maximizes the above likelihood function. Namely,

$$\hat{k}_s = \arg\max_{k_s}\{L\}. \tag{10}$$

To find $\hat{k}_s$, we take logarithm on both sides, and then perform the first order derivative to obtain

$$\frac{\partial \ln(L)}{\partial k_s} = \left(\frac{U_s^l}{q(k_s)} - \frac{l_s - U_s^l}{1-q(k_s)}\right) \times q'(k_s), \tag{11}$$

where $q'(k_s)$ is computed as

$$q'(k_s) = q(k_s) \times \ln\left(\frac{1-1/l_s}{1-1/m_s}\right). \tag{12}$$

Since $m_s > l_s \geq 1$ and $n_s > 0$, $q(k_s)$ and $q'(k_s)$ cannot be 0. Setting the right side of (11) be zero, we have

$$q(k_s) = \frac{U_s^l}{l_s} = V_s^l. \tag{13}$$

Substituting above equation to (8) and solving for $k_s$, we get the MLE estimator of $k_s$:

$$\hat{k}_s = \frac{\ln V_s^l - \ln V_s^m}{\ln(1-\frac{1}{l_s}) - \ln(1-\frac{1}{m_s})}. \tag{14}$$

Similarly, we can derive the MLE estimator of $k_r$:

$$\hat{k}_r = \frac{\ln V_r^l - \ln V_r^m}{\ln(1-\frac{1}{l_r}) - \ln(1-\frac{1}{m_r})}. \tag{15}$$

Since $k = k_s - k_r$, given the MLE estimators $\hat{k}_s$ and $\hat{k}_r$ of $k_s$ and $k_r$, we can easily derive the estimator of $k$ as

$$\hat{k} = \hat{k}_s - \hat{k}_r. \tag{16}$$

Substituting (14) and (15) to the above equation, we derive the estimator $\hat{k}$ as described in (4). Note that if the two bitmaps $B_s$ and $B_r$ have the same size, and the two logical bitmaps for each source host also have the same size, i.e., $m_s = m_r = m$ and $l_s = l_r = l$, then the estimator for the connection failure rate $k$ will be in a more compact form:

$$\hat{k} = \frac{\ln V_s^l - \ln V_s^m - \ln V_r^l + \ln V_r^m}{\ln(1-\frac{1}{l}) - \ln(1-\frac{1}{m})}. \tag{17}$$

## 4. A REGISTER ARRAY SOLUTION FOR LIMITING WORM PROPAGATION

In the double-bitmap solution, we discover the bitmap data structure has a limitation: it is difficult to extend the estimation range in such memory constrain without causing estimation inaccuracy. In order to address this problem, we propose another solution based on double shared register arrays called DSRA. Instead of sharing at bit level, DSRA tries to share at the register level. The size of each register is determined by the maximum estimation cardinality. For example, we can set the size of register to be 5 bits for the estimation up to $2^{32}$. We will incorporate two shared register arrays to store the SYN/SYN-ACK information for all source hosts. Our DSRA solution also includes two phases: SYN/SYN-ACK encoding and failure rate measurement. We will explain the two phases in detail, and then derive an estimator of the connection failure rate mathematically.

### 4.1. Phase I: SYN / SYN-ACK Encoding

In this phase, each edge router maintains two register arrays $M_s$ and $M_r$ to encode the distinct SYN packets and SYN-ACK packets of all source hosts. Let $t_s$ and $t_r$ be the number of registers in $M_s$ and $M_r$ correspondingly. In the register array $M_s$, the $i$th register is denoted by $M_s[i], 0 \leq i < t_s$. Next we will explain how an edge router encodes the distinct SYN packet information into $M_s$. The way for the router to encode the distinct SYN-ACK packet information into $M_r$ is quite similar.

Similar to our double-bitmap solution, in our DSRA solution, the edge router randomly selects $f_s$ ($\ll t_s$) registers from the physical register array $M_s$ to form a virtual register array for each source host $src$, which is denoted by $VR(src)$. The indices of the selected registers in $M_s$ are $H(src \oplus R[0])$, $H(src \oplus R[1])$, $\cdots$, $H(src \oplus R[f_s - 1])$, where $H(\cdots)$ is a hash function whose range is $[0, t_s)$. With these notations and data structures, the online coding works in the following. At the beginning of each measurement period, all registers in $M_s$ are reset to zeros. When a SYN packet signatured with a $\langle src, dst \rangle$ host address pair is routed by the edge router. The router will perform the hashing below:

$$p = H(src \oplus R[H(dst \oplus K) \bmod f_s])$$
$$q = H'(dst) = <x_1 x_2 ...>,$$

where $p$ is the hash result of $H(src \oplus R[H(dst \oplus K) \bmod f_s])$ and $q = <x_1 x_2 ...>$ is binary format of the hash output $H'(dst)$. Similar to online encoding phase of double-bitmap solution, the SYN packet is pseudo-randomly mapped to a register at index $p$. The operation to store the SYN packet information in this register is simple. Let $LZ(q)$ be the number of leading zeros in $q$ plus one. For example, if $q = 001...$, then $LZ(q) = 3$. We will update the mapped register if its current value is smaller than $LZ(q)$. Therefore, the overall effect to store the SYN packet information is:

$$M_s[p] = \max(M_s[p], LZ(q)).$$

The edge router only needs to set a value to a register in the register array for each SYN-ACK packet using the same mechanism. In addition, duplicates of SYN or SYN-ACK information with same $\langle src, dst \rangle$ signature will map to the same register and set the same value in the shared register array such that the duplicate information is filtered.

### 4.2. Phase II: Failure Rate Measurement

At the end of each measurement period, the edge router will send the two register arrays $M_s$ and $M_r$ to the NMC, which will estimate connection failure rate $k$ for each source host $src$ based on $M_s$ and $M_r$, and notify the edge router to apply rate limit algorithms if needed. The measurement process is described in the following.

First, the NMC extracts the virtual register arrays $VR(src)$ and $VR'(src)$ of each source host $src$ from the two shared register arrays $M_s$ and $M_r$, respectively. Then the NMC uses the HyperLogLog algorithm [43] to estimate the number of distinct elements in $M_s$, $M_r$, $VR(src)$ and $VR'(src)$, which are denoted by $\hat{n}_s$, $\hat{n}_r$, $\hat{n}_s^a$ and $\hat{n}_r^a$, respectively. Note that HyperLogLog algorithm is used to estimate the total number of distinct elements $\hat{n}$ that has been recorded by a register array $M$ with $f$ registers, denoted by $M[i], 0 \leq i < f$. The estimation from all registers in $M$ is

$$\hat{n} = \alpha_f \cdot f^2 \cdot \left( \sum_{j=0}^{f-1} 2^{-M[j]} \right)^{-1}, \tag{18}$$

where $\alpha_s$ is a constant which equals

$$\alpha_f = \left( f \int_0^\infty \left( \log_2(\frac{2+u}{1+u}) \right)^f du \right)^{-1}. \tag{19}$$

Finally, the NMC uses the following formula to estimate connection failure rate $k$ for source host $src$:

$$\hat{k} = \frac{t_s f_s}{t_s - f_s} \left( \frac{\hat{n}_s^a}{f_s} - \frac{\hat{n}_s}{t_s} \right) - \frac{t_r f_r}{t_r - f_r} \left( \frac{\hat{n}_r^a}{f_r} - \frac{\hat{n}_r}{t_r} \right) \tag{20}$$

### 4.3. Derivation of the estimator

We first use noise estimation method to get the estimators $\hat{k}_s$ and $\hat{k}_r$ of $k_s$ and $k_r$, respectively, and then derive $\hat{k}$ given by (20). Similarly, we will first derive the estimator formula for $\hat{k}_s$, and directly give the result for $\hat{k}_r$.

Recall that $n_s$ is the total number of all distinct SYN packets and $n_s^a$ is the number of distinct elements recorded by $VR(src)$, which is the source host $src$'s SYN packet cardinality plus the noise introduced by other source hosts who are sharing these registers. So the noise term in $VR(src)$ is $n_s^a - k_s$. Moreover, from the source host $src$'s view, the SYN packets of all other source hosts $(n_s - k_s)$ are noise. Since each noise packet has approximately an equal probability to be recorded by any register due to the random selection of registers by virtual register array, the average noise elements stored by an arbitrary register are $\dfrac{n_s - k_s}{t_s}$.

Hence, the total noise in the $f_s$ registers of $VR(src)$, $n_s^a - k_s$, can also be considered as the sum of $f_s$ independent random noise stored in the virtual register array:

$$E(n_s^a - k_s) = f_s \frac{n_s - k_s}{t_s}. \qquad (21)$$

By the law of large numbers in the probability theory, when $f_s$ is large, $E(n_s^a - k_s)$ can be replaced by an instance value, $n_s^a - k_s$. Hence, we have

$$n_s^a - k_s \approx f_s \frac{n_s - k_s}{t_s} \Rightarrow k_s \approx \frac{t_s f_s}{t_s - f_s}\left(\frac{n_s^a}{f_s} - \frac{n_s}{t_s}\right). \qquad (22)$$

Then, we get the estimator of $k_s$:

$$\hat{k}_s = \frac{t_s f_s}{t_s - f_s}\left(\frac{\hat{n}_s^a}{f_s} - \frac{\hat{n}_s}{t_s}\right). \qquad (23)$$

Similarly, we can derive the estimator of $k_r$:

$$\hat{k}_r = \frac{t_r f_r}{t_r - f_r}\left(\frac{\hat{n}_r^a}{f_r} - \frac{\hat{n}_r}{t_r}\right). \qquad (24)$$

Since $\hat{k} = \hat{k}_s - \hat{k}_r$, we can derive the estimator $\hat{k}$ as described in (20). Note that if $t_s = t_r = t$ and $f_s = f_r = f$, then the estimator for the connection failure rate $k$ can be presented in a more compact format:

$$\hat{k} = \frac{tf}{t-f}\left(\frac{\hat{n}_s^a - \hat{n}_r^a}{f} - \frac{\hat{n}_s - \hat{n}_r}{t}\right). \qquad (25)$$

# 5. SIMULATION

We evaluate the measurement accuracy of our estimator for the connection failure rate through simulations. Recall that the major goal of this paper is to provide a good estimator for measuring the connection failure rates of individual hosts that can work well in a small memory. Hence, in our simulations, we purposely allocate memory with small sizes to encode the information of distinct SYN and SYN-ACK packets for all source hosts, such that the average memory size for each source host will be ranging from 10 bits to 40 bits for double-bitmap solution, and ranging from 2.5 bits to 10 bits for DSRA solution. As we explained in Section 3.2, the solution with per-source address lists or bitmaps will not work with this small memory size. Therefore, our solutions outperform in the aspect of greatly reducing the required online memory footprint for connection failure rate measurement while achieving duplicate failure removal.

## 5.1. Simulation for double-bitmap solution

Our simulations are conducted under the following setups. We simulate 50,000 distinct source hosts as normal hosts, and 100 distinct source hosts as worm-affected hosts. For the normal hosts, they will send distinct SYN packets to different destination hosts, with a rate following an exponential distribution whose mean is 5 distinct SYN packets per minute. For each distinct SYN packet that a normal host sends out, a corresponding SYN-ACK packet will be sent back to the host with a probability, which follows a uniform distribution in the range of [0.8, 1.0]. As for the worm-affected hosts, we simulate their aggressive scanning behavior by having them send distinct SYN packets to different destination hosts with a higher rate, which follows another exponential distribution whose mean is 10 distinct SYN packets per second. Since the worm-affected hosts will randomly scan the whole destination space, their failure rate is expected to be very high as we explained earlier. Therefore, in our simulations, no SYN-ACK packets will be sent back to them. Suppose each measurement period is 1 minute. Then each normal host will send 5 distinct SYN packets and receive 4.5 distinct SYN-ACK packets on average, and each worm-affected host will send 600 distinct SYN packets and 0 SYN-ACK packet on average, during each measurement period.

In our simulations, all the SYN and SYN-ACK packets are processed by a single simulated edge router and a simulated network management center according to our two-phase measurement scheme. First of all, the SYN and SYN-ACK packets are encoded into two $m$-bit bitmaps $B_s$ and $B_r$ of the edge router, respectively, as described in Section 3.3.1 (Phase I: SYN/SYN-ACK Encoding). After all packets are encoded into the two bitmaps $B_s$ and $B_r$, the edge router will send $B_s$ and $B_r$ to the network management center, which will estimate the connection failure rate of each source host based on $B_s$ and $B_r$ offline, as described in Section 3.3.2 (Phase II: Failure Rate Measurement).

We conduct three sets of simulations with three different sizes of memory allocated for the bitmaps $B_s$ and $B_r$, $m_s = m_r = m =$ 0.5Mb, 1Mb, and 2Mb, to observe the measurement accuracy under different memory constraints. The sizes of the logical bitmaps for each host is set to be $l_s = l_r = l = 300$. Figure. 1-3 present the simulation results when the allocated memory $m$ equals 2Mb, 1Mb, and 0.5Mb, respectively. Since there are a total of 50,100 source hosts, the average memory consumption per source host will be about 40 bits, 20 bits, and 10 bits, accordingly. In each figure, each point represents a source host, with its x-coordinate showing the actual connection failure rate $k$ (per minute) and y-coordinate showing the estimated connection failure rate $\hat{k}$ (per minute) measured by our scheme. The equality line $y =$

*x* is also drawn for reference. Clearly, the closer a point is to the quality line, the better the measurement result.

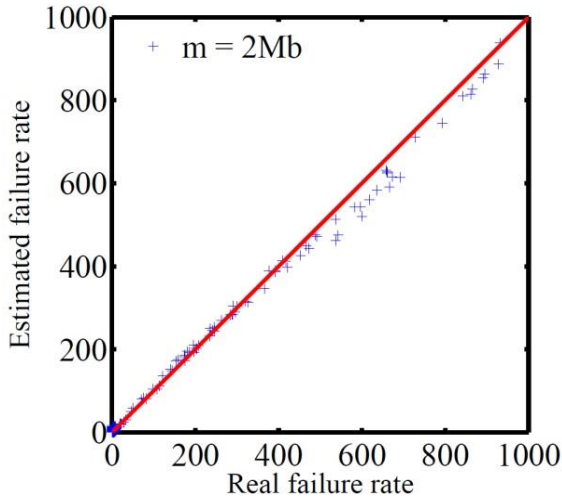

Figure 1. Measurement accuracy of connection failure rate per minute. $m = 2Mb$, $l = 300$.

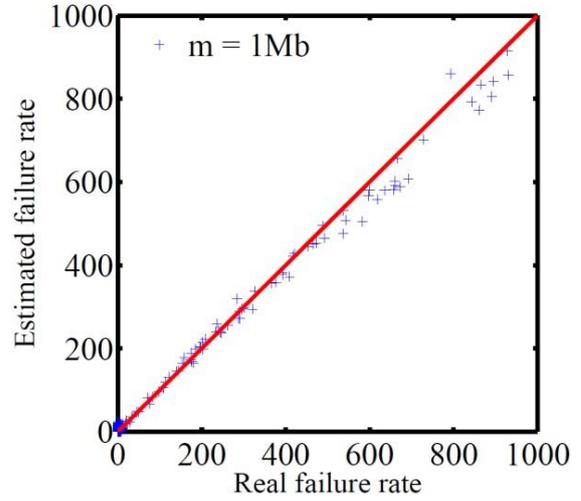

Figure 2. Measurement accuracy of connection failure rate per minute. $m = 1Mb$, $l = 300$.

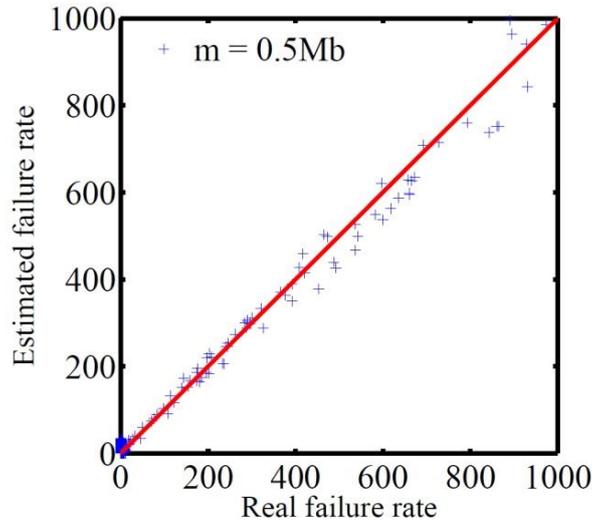

Figure 3. Measurement accuracy of connection failure rate per minute. $m = 0.5Mb$, $l = 300$.

From the three figures, one can observe that the measurement result for the connection failure rates of our scheme is quite accurate under all three different memory constraints. For almost every source host, the measured failure rate closely follows its real failure rate as shown in the figures. There is a tendency for the measurement result to be slightly more accurate with a larger memory size (compare Figure. 1 and Figure. 3). However, for our scheme, a small memory of size $m = 0.5Mb$ (equivalent to 10 bits per source host on average) is adequate enough to generate sound measurement results as shown in Figure. 3. Recall that for the solution storing per-source address list, the destination address of every SYN packet must be stored for every source host. So for that solution, a normal source host initiating 5 connection requests (5 distinct SYN packets) per minute will require at least $32 \times 5 = 160$ bits to record its SYN packets, and a worm-affected host sending 10 SYN packets per second will require at least $32 \times 600 = 19200$ bits, for each measurement period of one minute. Clearly, through utilizing double bitmaps, our scheme outperforms the solution storing address lists, because it can work well with a much more strict memory constraint.

## 5.2. Simulation for DSRA solution

We will evaluate the impact of memory space on the accuracy of connection failure rate estimation for double-bitmap solution and DSRA solution. For both solutions, we simulate 500,000 distinct source hosts as normal hosts, and 1,000 distinct source hosts as worm-affected hosts. The normal hosts will send distinct SYN packets to different destination hosts, with a rate following an exponential distribution whose mean is 5 distinct SYN packets per minute. The worm-affected hosts will send distinct SYN packets to different destination hosts with a higher rate, which follows another exponential distribution whose mean is 6000 distinct SYN packets per minute. The probability of sending back the SYC-ACK is the same as previous simulation. Suppose the measurement period is 1 minute, each normal host will send 5 distinct SYN packets and receive 4.5 distinct SYN-ACK packets on average, and each worm-affected host will send 6000 distinct SYN packets and 0 SYN-ACK packet on average.

To make a fair comparison, both solution will have the same size of memory to process the online encoding phase. We conduct three sets of simulations with three different sizes of memory allocated for the bitmaps $B_s$ and $B_r$, and the register arrays $M_s$ and $M_r$, $m_s = m_r = m = 5t_s = 5t_r = 5t = 5\text{Mb}$, $2.5\text{Mb}$, and $1.25\text{Mb}$, such that the average memory consumption per source host will be about 10 bits, 5 bits, and 2.5 bits, accordingly. We set $l_s = l_r = l = f_s = f_r = f = 512$. Figure. 4-6 present the simulation results when the allocate memory $m$ equals 5Mb, 2.5Mb, and 1.25Mb, respectively. Again, each point in each figure represents a source host, with its x-coordinate showing the actual connection failure rate $k$ (per minute) and y-coordinate showing the estimated connection failure rate $\hat{k}$ (per minute) measured by our schemes. The equality line $y = x$ is also drawn for reference.

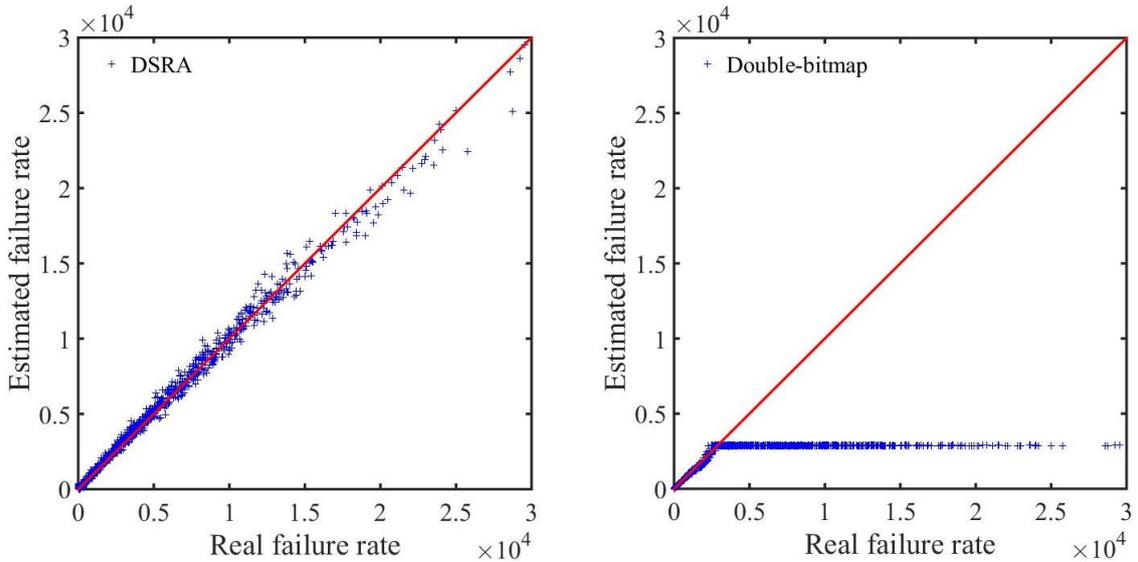

Figure 4: Compare DSRA and double-bitmap with 5Mb memory size

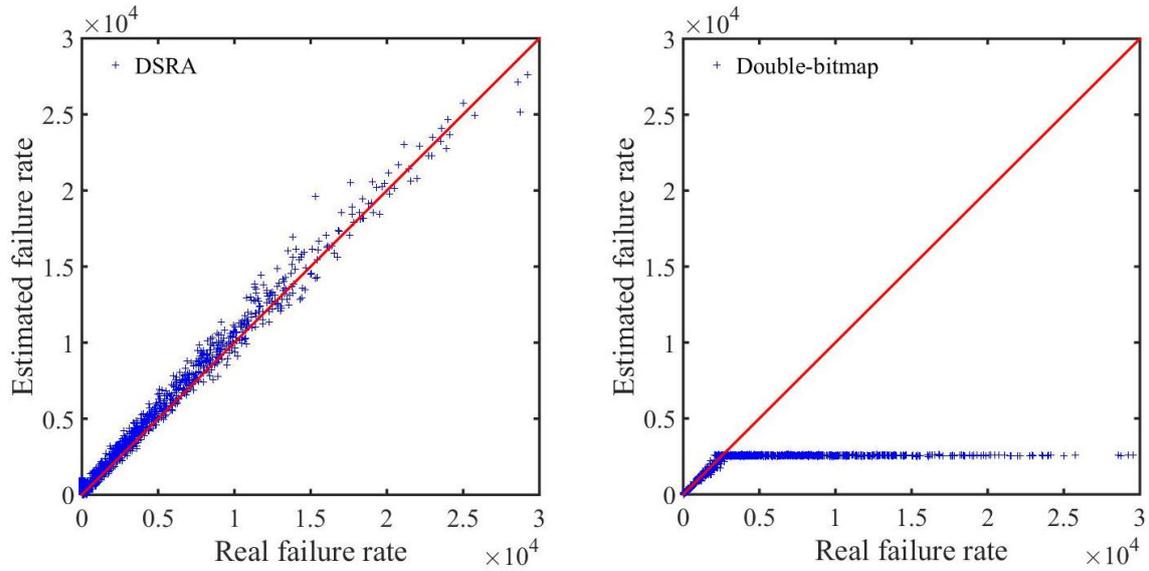

Figure 5: Compare DSRA and double-bitmap with 2.5Mb memory size

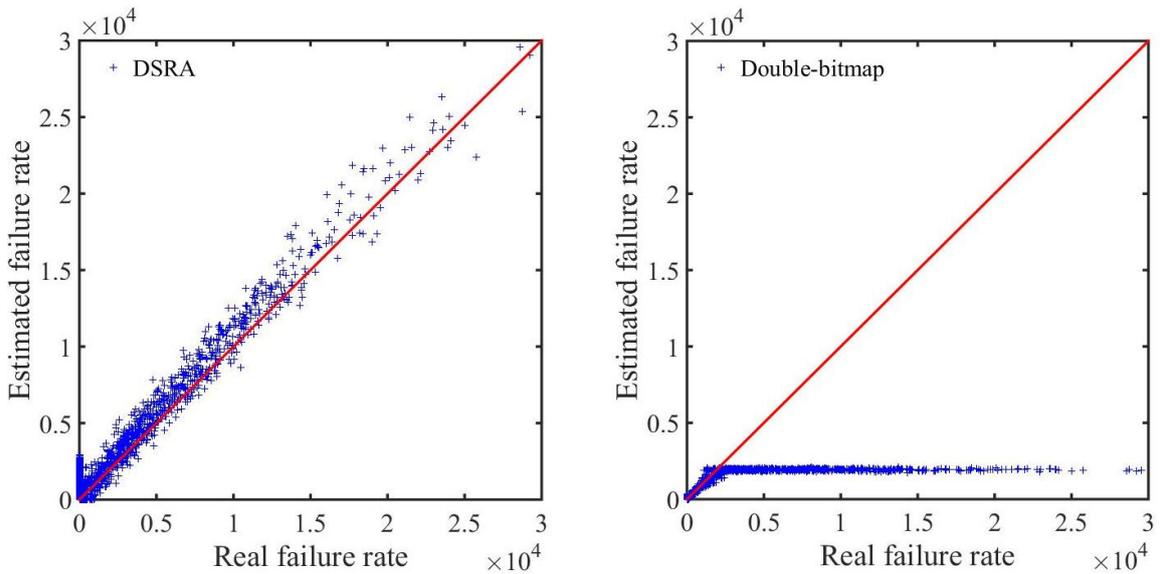

Figure 6: Compare DSRA and double-bitmap with 1.25Mb memory size

From the three figures, we can observe that the points in our DSRA scheme are clustered around the equality line. Therefore, the measurement result for the connection failure rates of our DSRA scheme is accurate under all three different memory constraints. As the memory size decreases, Figure. 4-6 show that the measurement result to be slightly deteriorating accuracy. However, the double-bitmap scheme is no more accurate when the actual connection failure rates are larger than around 2,500, which means its estimation upper bound is limited. One can observe that there is a tendency for the measurement upper bound for double-bitmap solution to be larger with a bigger memory size. Furthermore, for our DSRA scheme, small memory of size (equivalent to 2.5 bits per source host on average) is adequate enough to get sound measurement results as shown in Figure. 6. And the upper bound of DSRA scheme is $2^{32}$ which is appropriate in many real applications. Clearly, through above analyses, our DSRA scheme outperforms the double-bitmap scheme, because it can work well with a much more strict memory constraint and much larger estimation upper bound.

## 6. CONCLUSION

This paper proposes two new method of measuring connection failure rates of individual hosts, using two novel data structures: double bitmaps and double register arrays. It addresses an important problem in rate-limiting worm propagation, where inaccurate failure rates will affect the performance of rate-limit algorithms. The past method relies on ICMP host-unreachable messages, which are however widely blocked on today's Internet. The new methods make the measurement based on SYN and SYN-ACK packets, which are more reliable and accurate. The bitmap design helps significantly to reduce the memory footprint on the routers and eliminates the duplicate connection failures (another problem of the previous method). The register array design preserves the property of removing the duplicate connection failures, and achieves better memory efficiency and much larger estimation range than our bitmap solution.

## ACKNOWLEDGEMENTS

This work is supported in part by a grant from Florida Cybersecurity Center.

**AUTHORS**

**You Zhou** received his B.S. degree in electronic information engineering from the University of Science and Technology of China, Hefei, China, in 2013, and is currently pursuing his Ph.D. degree in computer and information science and engineering at the University of Florida, Gainesville, FL, USA. His advisor is Prof. Shigang Chen. His research interests include network security and privacy, big network data, and Internet of Things.

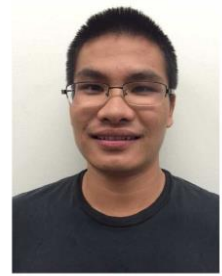

**Yian Zhou** received her B.S. degree in computer science and B.S. degree in economics from the Peking University of China in 2010, and her Ph.D. degree in computer and information science and engineering at the University of Florida, Gainesville, FL, USA in 2015. She is working with Google, Inc., Mountain View, CA, USA. Her research interests include traffic flow measurement, cyber-physical systems, big network data, security and privacy, and cloud computing.

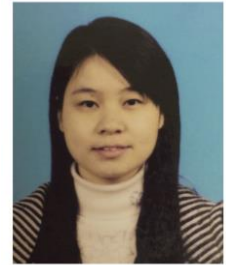

**Shigang Chen** is a professor with Department of Computer and Information Science and Engineering at University of Florida. He received his B.S. degree in computer science from University of Science and Technology of China in 1993. He received M.S. and Ph.D. degrees in computer science from University of Illinois at Urbana-Champaign in 1996 and 1999, respectively. After graduation, he had worked with Cisco Systems for three years before joining University of Florida in 2002. He served on the technical advisory board for Protego Networks in 2002-2003. His research interests include computer networks, Internet security, wireless communications, and distributed computing. He published more than 140 peer-reviewed journal/conference papers. He received IEEE Communications Society Best Tutorial Paper Award in 1999 and NSF CAREER Award in 2007. He holds 11 US patents. He is an associate editor for IEEE/ACM Transactions on Networking, Elsevier Journal of Computer Networks, and IEEE Transactions on Vehicular Technology. He served in the steering committee of IEEE IWQoS from 2010 to 2013. He is an IEEE fellow.

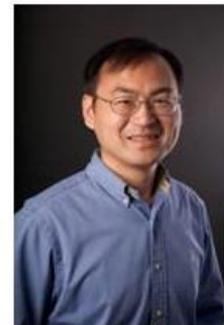

**O. Patrick Kreidl** has been an Assistant Professor of Electrical Engineering at the University of North Florida (UNF) since 2011, receiving his S.B. degree (with highest distinction) from George Mason University (GMU) in 1994 and his S.M. and Ph.D. degrees from the Massachusetts Institute of Technology (MIT) in 1996 and 2008, respectively. Past positions include Principal Research Engineer in the Cyber Operations and Networking Group within BAE Systems' Technology Solutions Directorate (via acquisition of Alphatech, Inc.), Research Affiliate in MIT's Laboratory for Information and Decision Systems, Adjunct Professor in GMU's Department of Electrical & Computer Engineering as well as engineering positions in the Institute for Defense Analyses and the Naval Research Laboratory. His current research interests lie at the intersections of signal processing, stochastic control and optimization (particularly as they interface with algorithms, computation and statistics) with application to sensor networks, network security and distributed systems. He is a member of the IEEE.

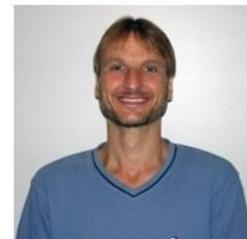